\documentclass[10pt]{extarticle}

\usepackage[bordercolor=white,backgroundcolor=gray!30,linecolor=black,colorinlistoftodos, textsize=10pt]{todonotes}

\usepackage{amssymb}

\usepackage{textcomp}
\usepackage{amsmath}
\usepackage[textstyle,amssymb]{SIunits}
\usepackage{graphicx}
\usepackage{fixltx2e}  
\usepackage[normalem]{ulem}  

\usepackage[twocolumn,textwidth=183mm,columnsep=6mm,top=20mm,bottom=25mm]{geometry}
\usepackage{helvet}
\usepackage{titlesec}
\titleformat*{\section}{\bfseries\sffamily}
\titlespacing{\section}{0pt}{*4}{*0}
\titleformat{\subsection}[runin]{\normalfont\bfseries}{\thesubsection.}{3pt}{}
\usepackage{setspace}
\usepackage[breaklinks=true]{hyperref} 

\usepackage[font={footnotesize,sf},labelfont={sf,bf},labelsep=endash,justification=raggedright]{caption}

\usepackage{natbib}
\usepackage[labelfont=bf,justification=justified]{caption}

\begin{document}

\twocolumn[\begin{@twocolumnfalse}
	{\huge\sf \textbf{Picosecond pulses from a mid-infrared interband cascade laser}}
	\vspace{0.5cm}
	
	{\sf\large \textbf {Johannes~Hillbrand$^{1,*}$, Maximilian~Beiser$^1$, Aaron~Maxwell~Andrews$^{1,2}$, Hermann~Detz$^{1,3}$, Robert~Weih$^4$, Anne~Schade$^5$, Sven~H\"ofling$^{5,6}$, Gottfried Strasser$^{1,2}$, Benedikt~Schwarz$^{1,+}$}}
		\vspace{0.5cm}
		
		{\sf \textbf{$^1$Institute of Solid State Electronics, TU Wien, Gusshausstraße 25-25a, 1040 Vienna, Austria\\
				$^2$Center for Micro- and Nanostructures, TU Wien, Gusshausstraße 25-25a, 1040 Vienna, Austria\\
				$^3$CEITEC, Brno University of Technology, Brno, Czech Republic\\
				$^4$Nanoplus Nanosystems and Technologies GmbH, 97218 Gerbrunn, Germany\\
				$^5$Technische Physik, Physikalisches Institut, University Würzburg, Würzburg, Germany\\
				$^6$SUPA, School of Physics and Astronomy, University of St Andrews, St Andrews, KY16 9SS, United Kingdom\\
				$^*$e-mail: {johannes.hillbrand@tuwien.ac.at}\\
				$^+$e-mail: {benedikt.schwarz@tuwien.ac.at}}}
		\vspace{0.5cm}
\end{@twocolumnfalse}]
\vspace{0.5cm}

{\sf \small \textbf{\boldmath
		\noindent The generation of mid-infrared pulses in monolithic and electrically pumped devices is of great interest for mobile spectroscopic instruments. The gain dynamics of interband cascade lasers (ICL) are promising for mode-locked operation at low threshold currents. Here, we present conclusive evidence for the generation of picosecond pulses in ICLs via active mode-locking. At small modulation power, the ICL operates in a linearly chirped frequency comb regime characterized by strong frequency modulation. Upon increasing the modulation amplitude, the chirp decreases until broad pulses are formed. Careful tuning of the modulation frequency minimizes the remaining chirp and leads to the generation of 3.2\,ps pulses.
	}
}

Optical frequency combs (OFC) operating in the mid-infrared (MIR) spectral range are a powerful spectroscopic tool~\cite{schliesser2012mid}. Measuring the molecular fingerprint in the MIR region allows to identify chemical species and to determine their concentration. MIR frequency comb sources based on the non-linear conversion of near-infrared mode-locked lasers~\cite{keilmann2012mid,leindecker2012octave} and microresonators~\cite{wang2013mid} have reached a high level of maturity featuring octave spanning spectra~\cite{kuyken2015octave}. Semiconductor laser OFCs~\cite{scalari2019onchip} are advantageous in applications requiring compactness and low power consumption.
Quantum cascade laser (QCL) frequency combs~\cite{faist1994quantum,hugi2012mid} are among the most investigated technologies. However, gain bandwidth and dispersion~\cite{villares2015dispersion,hillbrand2018tunable} limit their spectral bandwidth on the order of 100$\,$cm$^{-1}$. One way to overcome this issue is spectral broadening in an external non-linear fiber or waveguide. Recent results~\cite{singleton2018evidence} revealed, however, that the temporal output of QCLs is strongly chirped accompanied by the suppression of amplitude modulation. In fact, the ultrafast gain dynamics of QCLs are believed to be highly unfavorable for the formation of light pulses~\cite{gordon2008multimode}. Previous attempts of mode-locking in monolithic QCLs were limited to cryogenic temperatures and low peak powers~\cite{wang2009modelocked}. This makes non-linear techniques for spectral broadening very inefficient. First results to enter the mid-infrared from shorter wavelengths were demonstrated using passively mode-locked GaSb-based type-I cascade diode lasers with 10\,ps pulse duration around 3.25 \textmu m. 

\begin{figure}[b!]
	\centering
	\includegraphics[width=1\linewidth]{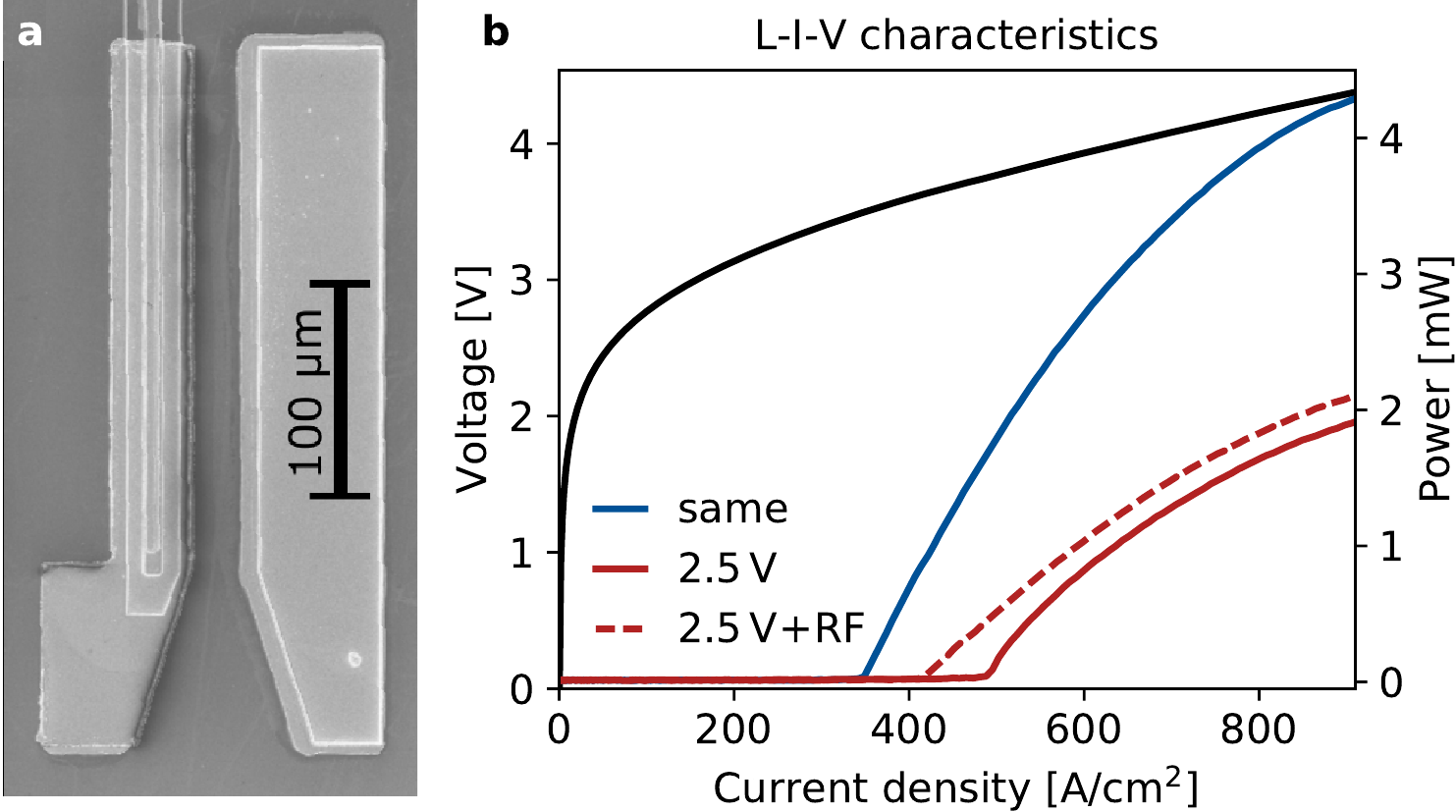}
	\caption{\textbf{a}: Scanning electron microscope picture of the modulation section of the ICL. The modulation section (left) and a ground contact (right) are optimized for RF injection via RF tips. \textbf{b}: Light-Current-Voltage (L-I-V) characteristics of the ICL at room temperature for a homogeneously biased laser (blue line) as well as for 2.5 V absorber bias with (dotted red line) and without (solid red line) RF modulation at $f_{rep}\approx 10.15\, $GHz. The RF power is 31 dBm.
	}
	\label{fig1}
\end{figure}

Type-II interband cascade lasers (ICL) are an interesting alternative that already cover a major part of the mid-infrared up to 6$\,\micro\meter$~\cite{yang1995infrared,vurgaftman2015interband,canedy2017interband}. They combine the carrier injection and extraction scheme of QCLs with the advantages of an interband lasing transition. Hence, the upper-state lifetime of the optical transition in ICLs is significantly longer than the cavity round-trip time. This enables low dissipation operation and has important consequences for mode-locking of ICLs. While the short upper-state lifetime of QCLs prevents the formation of short pulses, this issue is not present in ICLs.
Furthermore, the ICL active material can be switched to absorption at the laser wavelength, which was shown by using them as photodetector at zero-bias~\cite{lotfi2016monolithically}. Together with the fast carrier injection scheme, this allows the realization of efficient high-speed modulators with cut-off frequencies of several gigahertz~\cite{schwarz2018monolithic}.
Hence, ICLs exhibit all required properties for efficient active mode-locking via modulation of the gain at the cavity round-trip frequency $f_{rep}$~\cite{kuizenga1970fm}.
Recent efforts aimed at the generation of OFCs via passive mode-locking of ICLs. However, the experimental results did not show the formation of pulses~\cite{bagheri2018passively}. Instead, such passive ICL frequency combs are characterized by a continuous output intensity with a strong frequency modulation~\cite{schwarz2018monolithic}, similarly to what was found in QCLs~\cite{hugi2012mid,singleton2018evidence}.

In this letter, we report on the generation of picosecond pulses in two-section Fabry-P{\'e}rot ICLs.
The dry etched laser ridges are $6\,\micro\meter$ wide and split into a $3520\,\micro\meter$ long gain section and a $480\,\micro\meter$ long modulation section (Fig. \ref{fig1}a).
The modulation section was designed to minimize parasitic capacitance and allow efficient RF injection via coplanar RF tips.
A $1.5\,\micro\meter$ thick Si$_3$N$_4$ passivation layer was used for the modulation section while keeping its top contact area as small as possible. 
The passivation layer of the gain section is thinner ($250\,\nano\meter$) in order to improve the thermal performance of the laser. The back facet of the device was high-reflection coated using Si$_3$N$_4$ and gold, while the front facet was left uncoated. The active region is comprised of 6 stages and operates at 3.85$\,\micro\meter$ (2600$\,$cm$^{-1}$). At room temperature, the ICL emits up to $4.2\,\milli\watt$ of optical power in continuous wave operation when both sections are biased homogeneously (Fig. \ref{fig1}b).
When the bias of the modulation section is set to 2.5$\,$V additional loss is added to the cavity causing the threshold current density to increase and the maximum output power decreases to $1.9\,\milli\watt$.
The injection of an RF signal at $f_{rep}$ into the modulation section reduces the threshold by about 20\%, showing that the laser is strongly influenced by the active modulation.

\begin{figure}[b!]
	\centering
	\includegraphics[width=1\linewidth]{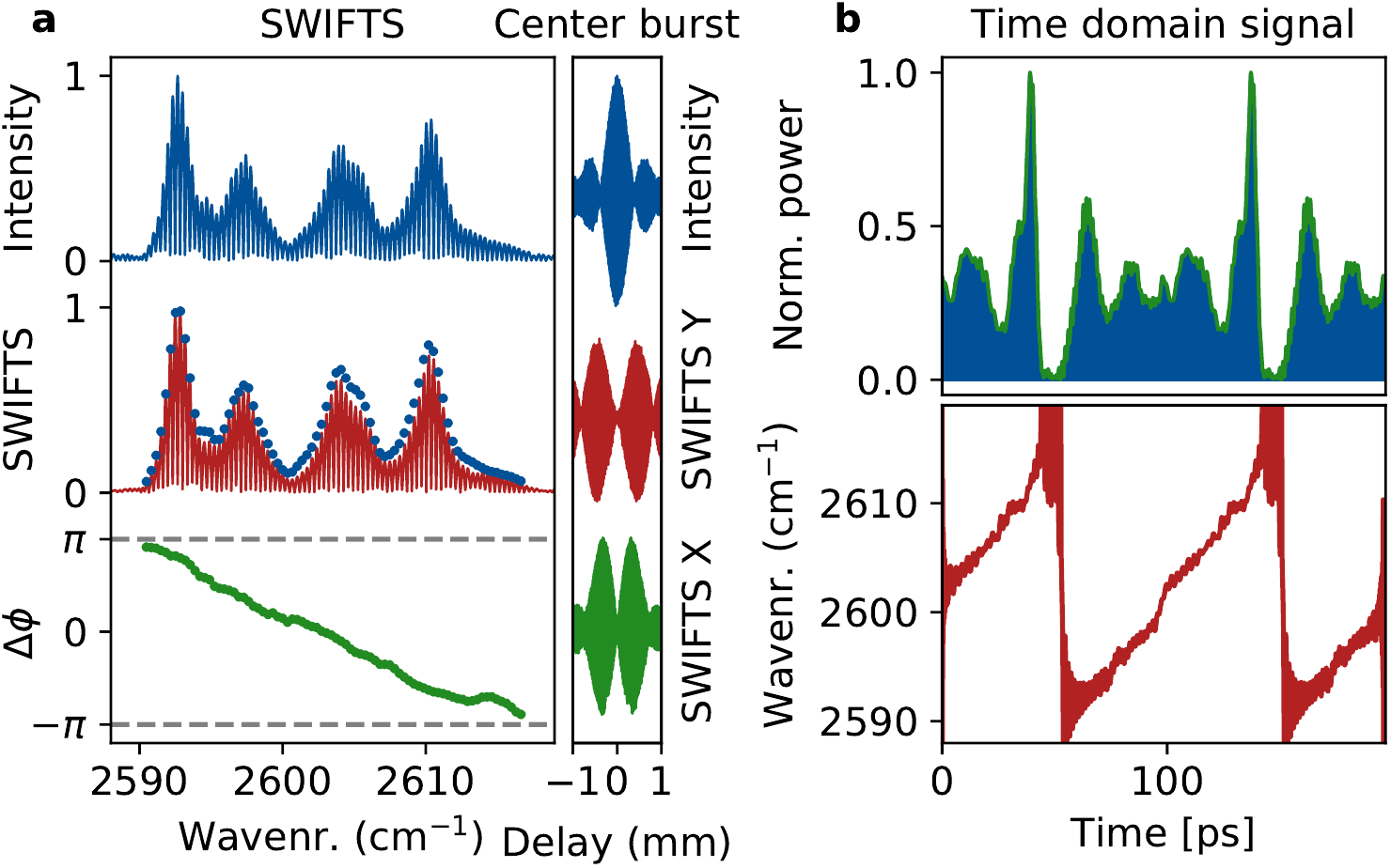}
	\caption{\textbf{a}: SWIFTS characterization of the ICL at -2 dBm injected RF power. The gain section is operated at 770$\,$A/cm$^2$ and the modulation section at 2.5$\,$V. blue line: intensity spectrum. red line: SWIFTS spectrum. blue dots: SWIFTS amplitudes expected for full phase-coherence of the ICL comb. green dots: intermodal difference phases of adjancent comb lines. The right part shows a zoom-in on the center burst of the intensity and SWIFTS interferograms. \textbf{b}: reconstructed intensity and instantaneous wavenumber of the ICL frequency comb.
	}
	\label{fig2}
\end{figure}
The characterization of the temporal output intensity of mid-infrared semiconductor lasers is challenging. Due to the high repetition rate and the relatively low average power, the peak power is expected to be too low for established non-linear pulse characterization techniques~\cite{kane1993characterization}.
Instead, we employ a linear phase-sensitive autocorrelation technique called 'SWIFTS'~\cite{burghoff2015evaluating}. This method uses a Fourier transform infrared (FTIR) spectrometer and a fast quantum well infrared photodetector (QWIP) to measure the amplitudes and phases of the beatings between adjacent laser modes (details can be found in Ref.~\cite{burghoff2015evaluating}).
In this way, SWIFTS allows the reconstruction of both the temporal intensity and instantaneous frequency of the ICL frequency comb. This method is not restricted to mode-locked operation and is valid for arbitrary periodic signals. Furthermore, it should be noted that recent experiments with a passively mode-locked quantum dot laser proved that SWIFTS and conventional intensity autocorrelation in a non-linear crystal are able to retrieve the same pulse width~\cite{hillbrand2019frequency}.
At 2.5$\,$V bias of the modulation section, the laser generates a narrow beatnote at the cavity round-trip frequency, which can be extracted directly from the laser current. This beatnote results from the beating of adjacent cavity modes. Its narrow linewidth on the kHz level indicates that the cavity modes are phase-locked. In order to provide a stable reference for SWIFTS, we inject a weak RF signal at -2$\,$dBm into the modulation section.
Previous experiments showed that such a weak modulation is able to lock the frequency of the beatnote and leaves the spectral phases of the free-running OFC unchanged~\cite{hillbrand2019coherent}.
The SWIFTS analysis of the ICL in this state is displayed in Fig. \ref{fig2}a. The intensity spectrum spans over roughly 28$\,$cm$^{-1}$ and consists of several lobes. The SWIFTS spectrum is commensurate with the values expected for full phase-coherence (blue dots in Fig. \ref{fig2}a) over the entire span of the spectrum, which proves frequency comb operation. The intermodal difference phases $\Delta \phi$ retrieved from the SWIFTS data decrease linearly over a range of exactly 2$\pi$. This particular frequency comb state was found in QCLs operating at 8$\,\micro\meter$~\cite{singleton2018evidence}, ICLs at 4$\,\micro\meter$~\cite{schwarz2018monolithic} and quantum dot lasers at 1.25$\,\micro\meter$~\cite{hillbrand2019frequency} and appears to be universal in semiconductor laser OFCs. Recent theoretical work attributes its origin to the interplay of dispersion and the Kerr effect in lasers with spatial hole burning~\cite{opacak2019theory}. Both SWIFTS interferograms (Fig. \ref{fig2}a right) have a local minimum at zero-path difference, which indicates the suppression of amplitude modulation~\cite{hugi2012mid}. Indeed, the reconstructed intensity (Fig. \ref{fig2}b top) does not show isolated pulses. In contrast, the instantaneous wavenumber is strongly modulated and linearly chirps through the entire spectrum within a cavity round-trip period.

\begin{figure*}
	\centering
	\includegraphics[width=0.91\linewidth]{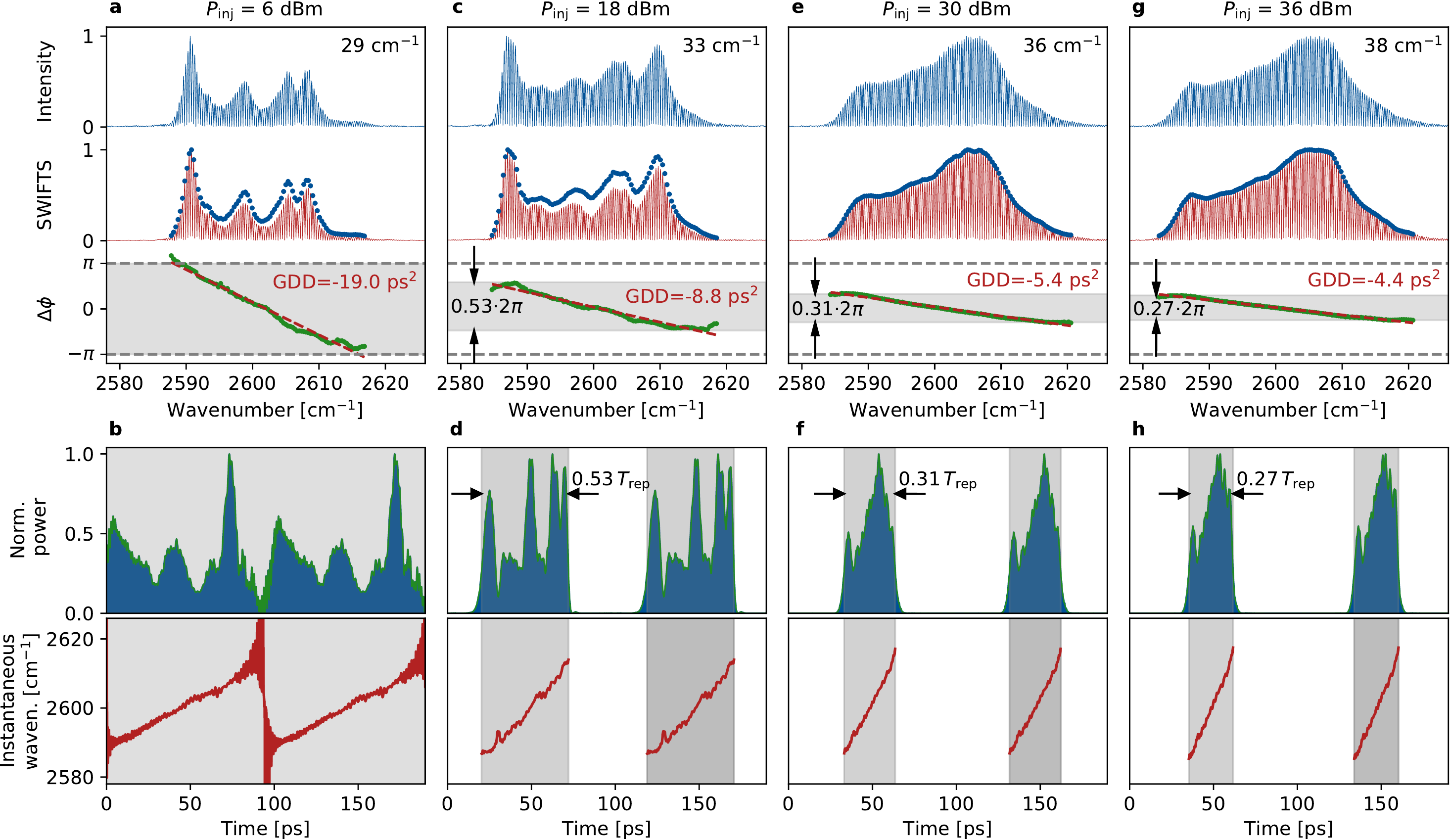}
	\caption{SWIFTS analysis of the ICL frequency comb for 6 dBm (\textbf{a}), 18 dBm (\textbf{c}), 30 dBm (\textbf{e}) and 36 dBm (\textbf{g}) injected power. The wavenumber in the top right corner indicates the bandwidth of the spectrum. The gain section was operated at 800 A/cm$^2$ and the modulation section bias was 2.5 V. The GDD is deduced from the slope of the intermodal difference phases (red dotted line). \textbf{b,d,f,h}: The corresponding reconstructed temporal intensity and instantaneous wavenumber. The shaded grey area indicates the fraction of phase range and of the cavity round-trip period, respectively, which is occupied by the pulses.
	}
	\label{fig3}
\end{figure*}

In the following, we will investigate the influence of an increased modulation strength on the ICL frequency comb dynamics. Fig. \ref{fig3} shows the detailed SWIFTS analysis of the ICL for injection power levels from 6$\,$dBm to 36$\,$dBm. The modulation frequency is altered slightly around $f_{rep}$ to account for a small detuning of the laser round-trip frequency with temperature due to the high injection power. At 6 dBm (Fig. \ref{fig3}a), the ICL still operates in a similar OFC state as in Fig. \ref{fig2}a and the reconstructed time signal (Fig. \ref{fig3}b) does not show isolated pulses. When the injected power is further increased to 18$\,$dBm (Fig. \ref{fig3}c), the spectral bandwidth grows to 33$\,$cm$^{-1}$ and the multiple lobes of the spectrum start to disappear. Interestingly, the SWIFTS spectrum shows that the ICL is not fully phase-locked in this transition state to the actively mode-locked regime. The intermodal difference phases still decrease linearly, but only cover the range of 0.53$\cdot 2 \pi$. Since $\Delta \phi$ is directly proportional to the group delay with 2$\pi$ corresponding to one cavity round-trip, this means that the ICL emits pulses with roughly 0.53$\cdot 2\pi$/2$\pi\approx$53\% duty cycle. Indeed, the reconstructed time signal (Fig. \ref{fig3}d) shows broad and linearly chirped pulses. At 30$\,$dBm, the spectrum consists of a single lobe spanning over 36$\,$cm$^{-1}$ (Fig. \ref{fig3}e). The slope of $\Delta \phi$ decreases, which corresponds to a decrease of the group delay dispersion (GDD) to -5.4$\,$ps$^2$ at 30$\,$dBm compared to -19$\,$ps$^2$ at 6$\,$dBm.
The reconstructed time signal (Fig. \ref{fig3}f) shows a train of isolated pulses with 27$\,$ps full width at half maximum (FWHM). When the injected power is increased by another 6$\,$dB, only a minor change in the shape of the spectrum and the GDD is observed (Fig. \ref{fig3}g). 

\begin{figure}[t!]
	\centering
	\includegraphics[width=0.9\linewidth]{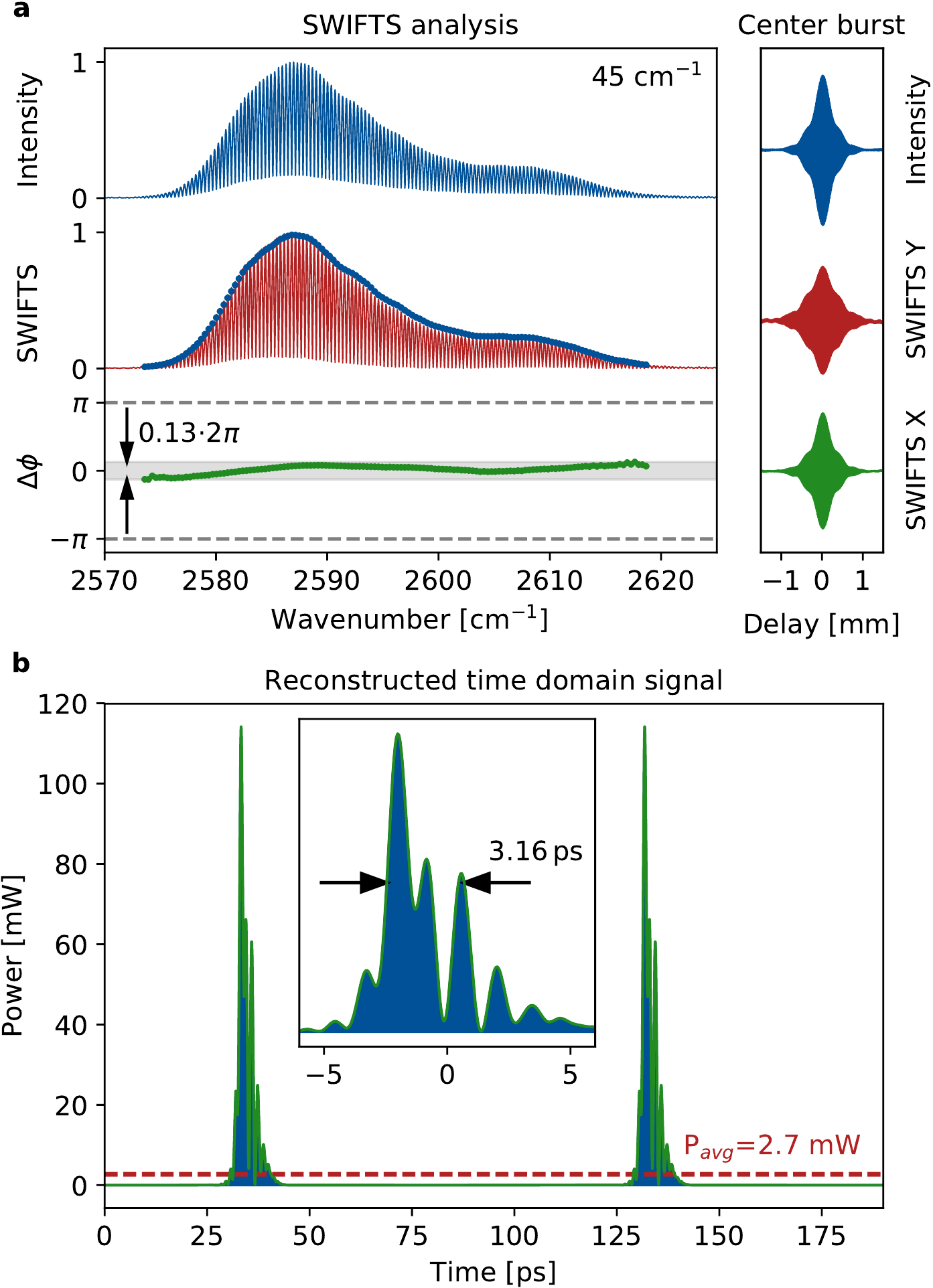}
	\caption{\textbf{a}: SWIFTS characterization for 920$\,$A/cm$^2$ gain section current density and 3 V modulation section bias. The modulation power is 32$\,$dBm and the modulation frequency is roughly 15$\,$MHz higher than $f_{rep}$. In contrast to the weakly modulated ICL (Fig. \ref{fig2}a), both SWIFTS interferograms now have local maxima at zero delay of the FTIR mirrors. This is an indicator for strong amplitude modulation of the laser intensity. \textbf{b}: Reconstructed time domain signal. The inset shows a zoom on a single pulse, revealing a FWHM of 3.2$\,$ps.
	}
	\label{fig4}
\end{figure}
Detuning the modulation frequency away from the round-trip frequency $f_{rep}$ of the free-running laser is another knob to influence the temporal output of the laser. Fig. \ref{fig4}a shows the SWIFTS characterization at 32$\,$dBm injected power, while the modulation frequency is 15$\,$MHz higher than $f_{rep}$. The spectrum consists of a single lobe spanning over 45\,$cm^{-1}$ and is fully coherent. Indeed, the intermodal difference phases do not show a linear chirp as in Figs \ref{fig3}a-h and occupy the phase range of only 0.13$\cdot 2\pi$. The reconstructed time signal (Fig. \ref{fig4}b) displays much shorter pulses than in Fig. \ref{fig3}g. A zoom on a single pulse reveals a FWHM of 3.2$\,$ps. The peak power is 114$\,$mW, which is an enhancement of more than 40 with respect to the average power of 2.7$\,$mW and proves that the energy can be efficiently stored by the gain medium over a round-trip. 
The inset in Fig.~\ref{fig4}b shows that the pulse is not transformation-limited with several smaller pulses arriving shortly after the first intense pulse. This is because the intermodal difference phases in Fig. \ref{fig4}a do not synchronize perfectly. 
In conclusion, our results show that ICLs provide all features necessary to generate MIR pulses within a single semiconductor laser ridge. The free-running or weakly modulated ICL frequency comb operates in a regime characterized by a strong linear chirp and suppression of amplitude modulation, which was also found in QCLs and quantum dot lasers. Upon increasing the modulation amplitude, the laser spectrum broadens accompanied by the formation of broad and chirped pulses. By carefully tuning the modulation frequency away from the round-trip frequency of the free-running laser, the pulse duration decreases to 3.2$\,$ps with a peak power enhancement of over 40. This illustrates the large potential of ICLs as a compact source for mid-infrared pulses.
From the spectral bandwidth one can assume that further optimization of the laser dispersion, modulation section length or adding additional segments will enable the generation of subpicosecond pulses~\cite{bowers1989actively}.

This work was supported by the Austrian Science Fund (FWF) within the projects "NanoPlas" (P28914-N27), "Building Solids for Function" (Project W1243), "NextLite" (F4909-N23), as well as by the Austrian Research Promotion Agency through the ERA-Net Photonic Sensing program, project "ATMO-SENSE" (FFG: 861581). J.H. was supported by the 'Hochschuljubil\"{a}umsstiftung' of the city of Vienna. H.D. was supported by the ESF project CZ.02.2.69/0.0/ 0.0/16\_027/0008371. A.M.A was supported by the projects COMTERA - FFG 849614 and AFOSR FA9550-17-1-0340. 
\bibliographystyle{naturemag}
\bibliography{literature}

\providecommand{\noopsort}[1]{}\providecommand{\singleletter}[1]{#1}%
\begin{thebibliography}{10}
\expandafter\ifx\csname url\endcsname\relax
  \def\url#1{\texttt{#1}}\fi
\expandafter\ifx\csname urlprefix\endcsname\relax\def\urlprefix{URL }\fi
\providecommand{\bibinfo}[2]{#2}
\providecommand{\eprint}[2][]{\url{#2}}

\bibitem{schliesser2012mid}
\bibinfo{author}{Schliesser, A.}, \bibinfo{author}{Picqu{\'e}, N.} \&
  \bibinfo{author}{H{\"a}nsch, T.~W.}
\newblock \bibinfo{title}{Mid-infrared frequency combs}.
\newblock \emph{\bibinfo{journal}{Nature Photonics}}
  \textbf{\bibinfo{volume}{6}}, \bibinfo{pages}{440--449}
  (\bibinfo{year}{2012}).

\bibitem{keilmann2012mid}
\bibinfo{author}{Keilmann, F.} \& \bibinfo{author}{Amarie, S.}
\newblock \bibinfo{title}{Mid-infrared frequency comb spanning an octave based
  on an er fiber laser and difference-frequency generation}.
\newblock \emph{\bibinfo{journal}{Journal of Infrared, Millimeter, and
  Terahertz Waves}} \textbf{\bibinfo{volume}{33}}, \bibinfo{pages}{479--484}
  (\bibinfo{year}{2012}).
\newblock \urlprefix\url{https://doi.org/10.1007/s10762-012-9894-x}.

\bibitem{leindecker2012octave}
\bibinfo{author}{Leindecker, N.} \emph{et~al.}
\newblock \bibinfo{title}{Octave-spanning ultrafast {OPO} with
  26-61$\mathrm{\mu}$m instantaneous bandwidth pumped by femtosecond tm-fiber
  laser}.
\newblock \emph{\bibinfo{journal}{Optics Express}}
  \textbf{\bibinfo{volume}{20}}, \bibinfo{pages}{7046} (\bibinfo{year}{2012}).
\newblock \urlprefix\url{https://doi.org/10.1364/oe.20.007046}.

\bibitem{wang2013mid}
\bibinfo{author}{Wang, C.~Y.} \emph{et~al.}
\newblock \bibinfo{title}{Mid-infrared optical frequency combs at
  2.5{\hspace{0.167em}}$\upmu$m based on crystalline microresonators}.
\newblock \emph{\bibinfo{journal}{Nature Communications}}
  \textbf{\bibinfo{volume}{4}} (\bibinfo{year}{2013}).
\newblock \urlprefix\url{https://doi.org/10.1038/ncomms2335}.

\bibitem{kuyken2015octave}
\bibinfo{author}{Kuyken, B.} \emph{et~al.}
\newblock \bibinfo{title}{An octave-spanning mid-infrared frequency comb
  generated in a silicon nanophotonic wire waveguide}.
\newblock \emph{\bibinfo{journal}{Nature Communications}}
  \textbf{\bibinfo{volume}{6}} (\bibinfo{year}{2015}).
\newblock \urlprefix\url{https://doi.org/10.1038/ncomms7310}.

\bibitem{scalari2019onchip}
\bibinfo{author}{Scalari, G.}, \bibinfo{author}{Faist, J.} \&
  \bibinfo{author}{Picqu{\'{e}}, N.}
\newblock \bibinfo{title}{On-chip mid-infrared and {THz} frequency combs for
  spectroscopy}.
\newblock \emph{\bibinfo{journal}{Applied Physics Letters}}
  \textbf{\bibinfo{volume}{114}}, \bibinfo{pages}{150401}
  (\bibinfo{year}{2019}).
\newblock \urlprefix\url{https://doi.org/10.1063/1.5097933}.

\bibitem{faist1994quantum}
\bibinfo{author}{Faist, J.} \emph{et~al.}
\newblock
  \bibinfo{title}{\href{https://doi.org/10.1126/science.264.5158.553}{Quantum
  Cascade Laser}}.
\newblock \emph{\bibinfo{journal}{Science}} \textbf{\bibinfo{volume}{264}},
  \bibinfo{pages}{553--556} (\bibinfo{year}{1994}).

\bibitem{hugi2012mid}
\bibinfo{author}{Hugi, A.}, \bibinfo{author}{Villares, G.},
  \bibinfo{author}{Blaser, S.}, \bibinfo{author}{Liu, H.~C.} \&
  \bibinfo{author}{Faist, J.}
\newblock \bibinfo{title}{Mid-infrared frequency comb based on a quantum
  cascade laser}.
\newblock \emph{\bibinfo{journal}{Nature}} \textbf{\bibinfo{volume}{492}},
  \bibinfo{pages}{229--233} (\bibinfo{year}{2012}).

\bibitem{villares2015dispersion}
\bibinfo{author}{Villares, G.} \emph{et~al.}
\newblock \bibinfo{title}{Dispersion engineering of quantum cascade laser
  frequency combs}.
\newblock \emph{\bibinfo{journal}{Optica}} \textbf{\bibinfo{volume}{3}},
  \bibinfo{pages}{252--258} (\bibinfo{year}{2016}).

\bibitem{hillbrand2018tunable}
\bibinfo{author}{Hillbrand, J.}, \bibinfo{author}{Jouy, P.},
  \bibinfo{author}{Beck, M.} \& \bibinfo{author}{Faist, J.}
\newblock \bibinfo{title}{Tunable dispersion compensation of quantum cascade
  laser frequency combs}.
\newblock \emph{\bibinfo{journal}{Optics Letters}}
  \textbf{\bibinfo{volume}{43}}, \bibinfo{pages}{1746} (\bibinfo{year}{2018}).
\newblock \urlprefix\url{https://doi.org/10.1364/ol.43.001746}.

\bibitem{singleton2018evidence}
\bibinfo{author}{Singleton, M.}, \bibinfo{author}{Jouy, P.},
  \bibinfo{author}{Beck, M.} \& \bibinfo{author}{Faist, J.}
\newblock \bibinfo{title}{Evidence of linear chirp in mid-infrared quantum
  cascade lasers}.
\newblock \emph{\bibinfo{journal}{Optica}} \textbf{\bibinfo{volume}{5}},
  \bibinfo{pages}{948} (\bibinfo{year}{2018}).
\newblock \urlprefix\url{https://doi.org/10.1364/optica.5.000948}.

\bibitem{gordon2008multimode}
\bibinfo{author}{Gordon, A.} \emph{et~al.}
\newblock \bibinfo{title}{Multimode regimes in quantum cascade lasers: From
  coherent instabilities to spatial hole burning}.
\newblock \emph{\bibinfo{journal}{Physical Review A}}
  \textbf{\bibinfo{volume}{77}} (\bibinfo{year}{2008}).
\newblock \urlprefix\url{https://doi.org/10.1103/physreva.77.053804}.

\bibitem{wang2009modelocked}
\bibinfo{author}{Wang, C.~Y.} \emph{et~al.}
\newblock \bibinfo{title}{Mode-locked pulses from mid-infrared quantum cascade
  lasers}.
\newblock \emph{\bibinfo{journal}{Optics Express}}
  \textbf{\bibinfo{volume}{17}}, \bibinfo{pages}{12929} (\bibinfo{year}{2009}).
\newblock \urlprefix\url{https://doi.org/10.1364/oe.17.012929}.

\bibitem{yang1995infrared}
\bibinfo{author}{Yang, R.~Q.}
\newblock \bibinfo{title}{Infrared laser based on intersubband transitions in
  quantum wells}.
\newblock \emph{\bibinfo{journal}{Superlattices and Microstructures}}
  \textbf{\bibinfo{volume}{17}}, \bibinfo{pages}{77--83}
  (\bibinfo{year}{1995}).
\newblock \urlprefix\url{https://doi.org/10.1006/spmi.1995.1017}.

\bibitem{vurgaftman2015interband}
\bibinfo{author}{Vurgaftman, I.} \emph{et~al.}
\newblock \bibinfo{title}{Interband cascade lasers}.
\newblock \emph{\bibinfo{journal}{Journal of Physics D: Applied Physics}}
  \textbf{\bibinfo{volume}{48}}, \bibinfo{pages}{123001}
  (\bibinfo{year}{2015}).
\newblock \urlprefix\url{https://doi.org/10.1088/0022-3727/48/12/123001}.

\bibitem{canedy2017interband}
\bibinfo{author}{Canedy, C.~L.} \emph{et~al.}
\newblock \bibinfo{title}{Interband cascade lasers with longer wavelengths}.
\newblock In \bibinfo{editor}{Razeghi, M.} (ed.)
  \emph{\bibinfo{booktitle}{Quantum Sensing and Nano Electronics and Photonics
  {XIV}}} (\bibinfo{publisher}{{SPIE}}, \bibinfo{year}{2017}).
\newblock \urlprefix\url{https://doi.org/10.1117/12.2246450}.

\bibitem{lotfi2016monolithically}
\bibinfo{author}{Lotfi, H.} \emph{et~al.}
\newblock \bibinfo{title}{Monolithically integrated mid-{IR} interband cascade
  laser and photodetector operating at room temperature}.
\newblock \emph{\bibinfo{journal}{Applied Physics Letters}}
  \textbf{\bibinfo{volume}{109}}, \bibinfo{pages}{151111}
  (\bibinfo{year}{2016}).
\newblock \urlprefix\url{https://doi.org/10.1063/1.4964837}.

\bibitem{schwarz2018monolithic}
\bibinfo{author}{Schwarz, B.} \emph{et~al.}
\newblock \bibinfo{title}{A monolithic frequency comb platform based on
  interband cascade lasers and detectors}.
\newblock \emph{\bibinfo{journal}{arXiv preprint arXiv:1812.03879}}
  (\bibinfo{year}{2018}).

\bibitem{kuizenga1970fm}
\bibinfo{author}{Kuizenga, D.} \& \bibinfo{author}{Siegman, A.}
\newblock \bibinfo{title}{{FM} and {AM} mode locking of the homogeneous laser -
  part i: Theory}.
\newblock \emph{\bibinfo{journal}{{IEEE} Journal of Quantum Electronics}}
  \textbf{\bibinfo{volume}{6}}, \bibinfo{pages}{694--708}
  (\bibinfo{year}{1970}).
\newblock \urlprefix\url{https://doi.org/10.1109/jqe.1970.1076343}.

\bibitem{bagheri2018passively}
\bibinfo{author}{Bagheri, M.} \emph{et~al.}
\newblock \bibinfo{title}{Passively mode-locked interband cascade optical
  frequency combs}.
\newblock \emph{\bibinfo{journal}{Scientific Reports}}
  \textbf{\bibinfo{volume}{8}} (\bibinfo{year}{2018}).
\newblock \urlprefix\url{https://doi.org/10.1038/s41598-018-21504-9}.

\bibitem{kane1993characterization}
\bibinfo{author}{Kane, D.} \& \bibinfo{author}{Trebino, R.}
\newblock \bibinfo{title}{Characterization of arbitrary femtosecond pulses
  using frequency-resolved optical gating}.
\newblock \emph{\bibinfo{journal}{{IEEE} Journal of Quantum Electronics}}
  \textbf{\bibinfo{volume}{29}}, \bibinfo{pages}{571--579}
  (\bibinfo{year}{1993}).
\newblock \urlprefix\url{https://doi.org/10.1109/3.199311}.

\bibitem{burghoff2015evaluating}
\bibinfo{author}{Burghoff, D.} \emph{et~al.}
\newblock \bibinfo{title}{Evaluating the coherence and time-domain profile of
  quantum cascade laser frequency combs}.
\newblock \emph{\bibinfo{journal}{Optics Express}}
  \textbf{\bibinfo{volume}{23}}, \bibinfo{pages}{1190} (\bibinfo{year}{2015}).

\bibitem{hillbrand2019frequency}
\bibinfo{author}{Hillbrand, J.} \emph{et~al.}
\newblock \bibinfo{title}{Frequency comb dynamics of ultrafast quantum dot
  lasers}.
\newblock \emph{\bibinfo{journal}{currently being prepared}}
  (\bibinfo{year}{2019}).

\bibitem{hillbrand2019coherent}
\bibinfo{author}{Hillbrand, J.}, \bibinfo{author}{Andrews, A.~M.},
  \bibinfo{author}{Detz, H.}, \bibinfo{author}{Strasser, G.} \&
  \bibinfo{author}{Schwarz, B.}
\newblock \bibinfo{title}{Coherent injection locking of quantum cascade laser
  frequency combs}.
\newblock \emph{\bibinfo{journal}{Nature Photonics}}
  \textbf{\bibinfo{volume}{13}}, \bibinfo{pages}{101--104}
  (\bibinfo{year}{2018}).
\newblock \urlprefix\url{https://doi.org/10.1038/s41566-018-0320-3}.

\bibitem{opacak2019theory}
\bibinfo{author}{Opa{\v{c}}ak, N.} \& \bibinfo{author}{Schwarz, B.}
\newblock \bibinfo{title}{Theory of frequency modulated combs induced by
  spatial hole burning, dispersion and kerr}.
\newblock \emph{\bibinfo{journal}{arXiv preprint arXiv:1905.13635}}
  (\bibinfo{year}{2019}).

\bibitem{bowers1989actively}
\bibinfo{author}{Bowers, J.}, \bibinfo{author}{Morton, P.},
  \bibinfo{author}{Mar, A.} \& \bibinfo{author}{Corzine, S.}
\newblock \bibinfo{title}{Actively mode-locked semiconductor lasers}.
\newblock \emph{\bibinfo{journal}{{IEEE} Journal of Quantum Electronics}}
  \textbf{\bibinfo{volume}{25}}, \bibinfo{pages}{1426--1439}
  (\bibinfo{year}{1989}).
\newblock \urlprefix\url{https://doi.org/10.1109/3.29278}.

\end{thebibliography}


\end{document}